# Nodal-Surface and Flat-Band Driven Large Anomalous Nernst Effect in Epitaxial Ferromagnetic Weyl Metal $Fe_5Si_3$

Shubhashish Pati[1], Sonali Srotaswini Pradhan[2], Nanhe Kumar Gupta[3], Abhay Pandey[1], Nikita Sharma[1], Nakul Kumar[1], Saurav Singh[1], Yuya Sakuraba[3], V. Kanchana[2]*, Sujeet Chaudhary[1]*

[1] Thin Film Laboratory, Department of Physics, Indian Institute of Technology Delhi, New Delhi 110016, India

[2] Department of Physics, Indian Institute of Technology Hyderabad, Kandi, Sangareddy, Telangana 502285, India

[3] Centre for Magnetic and Spintronic Materials (CMSM), National Institute for Materials Science (NIMS) 1-2-1 Sengen, Tsukuba, Ibaraki 305-0047, Japan.

## Abstract

Magnetic topological materials such as Weyl and Dirac magnets exhibit unconventional electronic properties arising from the interplay between magnetic order and band topology, leading to notable thermomagnetic and thermoelectric effects. Unlike conventional Seebeck-based thermoelectric devices, the Nernst-based devices exploit the anomalous Nernst effect (ANE), where Berry curvature induces a transverse voltage in response to a thermal gradient, enabling efficient single-material-driven energy conversion. Here, we investigate the ANE in epitaxial thin films of the Weyl ferromagnet candidate $Fe_5Si_3$. A pronounced transverse Nernst response exceeding ~1.50 µV K$^{-1}$ is observed at room temperature, together with a giant anomalous Nernst angle of ~ 0.56, reflecting an efficient conversion between longitudinal thermal gradients and transverse electric fields. Beyond the anomalous contribution, a substantial topological Nernst signal of ~ 0.43 µV K$^{-1}$ persists above room temperature, indicating the possible existence of real-space Berry curvature arising from nontrivial spin-textures. First-principles density functional theory (DFT) calculations, combined with symmetry analysis, reveal an unconventional electronic structure in which Weyl nodal lines, nodal surfaces, and nearly flat bands coexist near the Fermi level. This rare concurrence of multiple topological band features generates a strongly enhanced and sharply energy-dependent Berry curvature, which dictates both the magnitude and temperature evolution of the measured Nernst response. The close quantitative agreement between the calculated anomalous Nernst conductivity and experimental results establishes the topological electronic structure as the primary origin of the observed thermomagnetic transport, highlighting $Fe_5Si_3$ as a unique, low-cost, and chemically simple binary topological magnet for probing real-space as well as reciprocal Berry-curvature-driven thermoelectric phenomena.

Corresponding Authors

E-mail address: sujeetc@physics.iitd.ac.in (S. Chaudhary), kanchana@phy.iith.ac.in (V. Kanchana)

## Introduction

The magnetic topological materials possess large Berry curvature, a fictitious magnetic field in momentum space that gives rise to intrinsic transverse transport phenomena such as the anomalous Hall effect (AHE) and its thermoelectric counterpart, the anomalous Nernst effect (ANE).[1–4] The ANE enables direct heat-to-electricity conversion by generating a voltage perpendicular to both the temperature gradient ($\nabla T$) and the magnetisation (M), offering advantages such as simplified lateral device architectures and efficient coverage of curved or large-area heat sources.[5–8] Unlike conventional thermoelectric effects dominated by magnetization, the ANE arises intrinsically from the Berry curvature near the Fermi level ($E_F$), making Topological materials, where time-reversal or inversion symmetry is broken and Weyl points or nodal lines emerge, an ideal platform for realizing enhanced performance[1,9]. Recent studies have demonstrated the potential of topological band engineering for realizing large ANE responses. Remarkably, materials such as the ferromagnetic Weyl semimetal $Co_2MnGa$ exhibit a giant ANE of ≈ 6 μV $K^{-1}$ at room temperature[5], while low-cost binary compounds like $Fe_3Ga$ and $Fe_3Al$ display strong ANE responses driven by dense nodal web structures near $E_F$[10], and the chiral antiferromagnet $Mn_3Sn$ achieves ≈0.6 μV $K^{-1}$ at 200 K despite negligible magnetization.[2] These advances underscore that tuning Berry curvature through topological band engineering provides a powerful strategy for developing next-generation, high-efficiency transverse thermoelectric materials for sustainable energy applications.

Building on the concepts of nodal-point and nodal-line topological materials, exploring higher-dimensional band crossings such as nodal chains, nodal web and nodal surfaces presents an exciting frontier for realizing enhanced topological responses. These extended nodal structures are expected to generate broader regions of intense Berry curvature near the Fermi level, potentially amplifying transverse transport phenomena. The reports on such systems remain exceedingly scarce, with $Fe_3Sn$[11] and $CoNb_3S_6$[12] being the only known examples where a nodal plane band topology has been theoretically established, accompanied by an experimental verification of significant anomalous Nernst response. The rarity of such studies highlights a significant opportunity to investigate multi-dimensional topological features as a new design principle for high-performance thermoelectric materials.

Here, we report the experimental observation of a pronounced ANE in the ferromagnetic Weyl metal candidate $Fe_5Si_3$. Through a combination of detailed experimental measurements and first-principles DFT calculations, we uncover the emergence of a nodal surface and a nearly flat band, coexisting with previously explored Weyl nodal line features. Despite its promising topological characteristics, $Fe_5Si_3$ has been unexplored from a thermoelectric perspective. Structurally, it crystallizes in a hexagonal $D8_8$-type lattice (space group $P6_3/mmc$), identical to that of other advanced transition metal-based topological materials ($A_5B_3$, where A = Fe, Co, Mn and B = Si, Sn, Ge ) known for exhibiting exceptionally large AHE and ANE responses[11,13–15]. Thermal transport measurements on 50 nm epitaxial $Fe_5Si_3$ thin films reveal a pronounced Nernst coefficient of ~ 1.50 µV K$^{-1}$ near room temperature, nearly five times greater than that of conventional ferromagnetic systems. The anomalous Nernst conductivity (ANC) attains 1.50 A m$^{-1}$ K$^{-1}$ at 150 K, showing excellent agreement with ab initio DFT predictions. Remarkably, a giant anomalous Nernst angle ($\theta_{ANE} \approx 0.56$) is observed at 175 K and remains significant ($\theta_{ANE} \approx 0.2$) at room temperature, demonstrating efficient thermoelectric conversion. This unusually large angle highlights the dominant role of Berry curvature near the Fermi level, resulting in a substantial transverse carrier deflection. In addition, a sizable topological Nernst resistivity of ~ 0.43 µV K$^{-1}$ persists above room temperature, suggesting the emergence of chiral spin textures within the system. This response highlights a real-space topological contribution linked to nontrivial spin chirality and finite scalar spin correlations. The close correspondence between experiment and theory confirms the intrinsic origin of the transverse thermoelectric response and firmly establishes $Fe_5Si_3$ as a promising, low-cost magnetic topological thermoelectric material for high-performance energy conversion applications.

## Experimental Details

The epitaxial $Fe_5Si_3$ thin films of thickness 50 nm were grown using a DC magnetron co-sputtering system. The detailed growth procedure and the DC magnetization characteristics of $Fe_5Si_3$ are reported elsewhere.[16] A comprehensive analysis of the AHE, including its intrinsic Berry-curvature driven origin and the associated scaling behaviour, has been presented.[16] For completeness, we also measured the AHE in the device-patterned geometry used for ANE measurements. Since the patterned and unpatterned films were prepared from the same batch of epitaxial $Fe_5Si_3$ films, the resulting hysteresis loops, as well as the extracted anomalous Hall conductivity values, closely match those obtained from unpatterned films (see Fig. S(1) of SI), confirming that device fabrication does not alter the underlying transport properties.

Thermoelectric measurements, including Seebeck (SE) and ANE characterizations, were performed on device-patterned $Fe_5Si_3$ thin films. Details of the device fabrication are provided in the section 1 of SI. Zero-field temperature-dependent Seebeck coefficients were measured over a range of 75–325 K with 25 K intervals. The longitudinal Seebeck coefficient ($S_{xx}$) was determined using $S_{xx} = \frac{V_x}{\nabla T_x}$, where $V_x$ is the longitudinal voltage and $\nabla T_x$ is the applied thermal gradient. ANE measurements were conducted in the 75-325 K range under magnetic fields up to ±3 T. The transverse Nernst thermopower ($S_{xy}^A$) was estimated using $S_{xy}^A = \frac{L_x V_y}{L_y \nabla T_x}$, where $V_y$ is the transverse voltage, $L_x$ is the distance between the two temperature leads, and $L_y$ is the distance between the two voltage probes.

## Results and Discussions

Figure 1 (d) presents the zero-field temperature dependence of $S_{xx}$, calculated from the slope of the linear fit between the electric field ($E_x$) generated by the SE and the applied thermal gradient $\nabla T_x$ (See sec.1 in SI). The $S_{xx}$ shows a negative sign throughout the measurement temperature ranges, indicative of electron-dominated thermoelectric transport. Furthermore, the magnitude of $S_{xx}$ increases monotonically with temperature, reaching approximately 5.63 µV K$^{-1}$ at room temperature. The continuous increase of $S_{xx}$ with temperature clearly indicates the absence of bipolar conduction effects, arising from simultaneous electron and hole contributions in the $Fe_5Si_3$ system.[5]

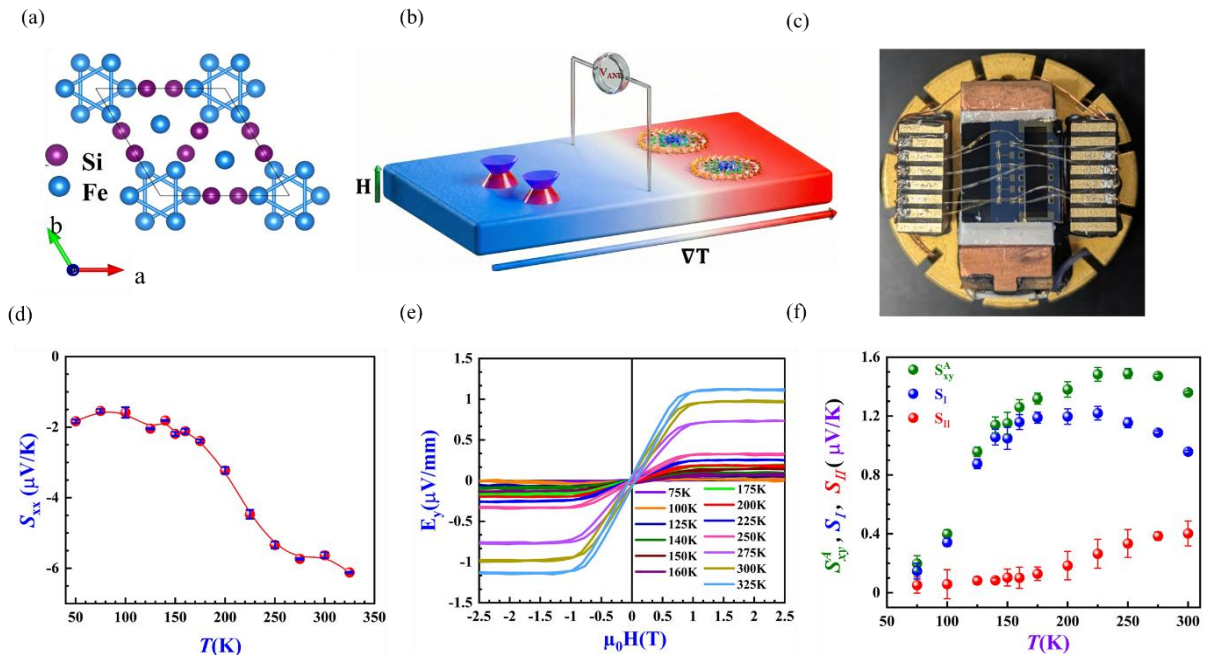

**Figure 1**. (a) Schematic view of the crystal structure of $Fe_5Si_3$. (b) A Schematic illustration of Transverse anomalous Nernst voltage generation in single material geometry. (c) Real image of patterned $Fe_5Si_3$ film and sample holder for ANE and SE measurements. (d) Temperature-dependent zero-field Seebeck coefficient ($S_{xx}$). (e) Magnetic field dependence of ANE-induced $E_y$ at different temperatures. (f) Temperature-dependent $S_{xy}^A$, and competing contributions $S_I$ and $S_{II}$.

The Nernst effect in epitaxial $Fe_5Si_3$ thin film was systematically investigated to elucidate the underlying mechanisms governing its anomalous transverse thermoelectric response. Figure 1(e) displays the magnetic field dependence of the ANE-induced transverse electric field ($E_y$) measured over a temperature range of 75-325K. The corresponding $S_{xy}^A$ values are extracted from the slope of $E_y$ with respect to the applied thermal gradient $\nabla T_x$. Across all temperature ranges, the Nernst signal increases rapidly at low magnetic field and saturates around 1.2T, which is consistent with out of plane (OOP) magnetization and AHE behaviour. With increasing temperature from 75K to 325K, the absolute value of the ANE coefficient increases monotonically and reaches a maximum value of 1.50 µV K⁻¹ at room temperature, comparable to the values reported for other state-of-the-art topological materials[2,10,17–19] and significantly larger than the conventional ferromagnets, underscoring unconventional thermoelectric behaviour of $Fe_5Si_3$. To gain deeper insight into the nature of the anomalous Nernst responses, we estimate the ANC, $\alpha_{xy}^A$ using the equation, $\alpha_{xy}^A = \sigma_{xx}S_{xy}^A + \sigma_{xy}S_{xx}$, where $\sigma_{xx}$ and $\sigma_{xy}$ represent the longitudinal and transverse electrical conductivities, respectively. In conventional ferromagnetic metals, $\alpha_{xy}^A$ typically decreases with increasing temperature, following the temperature dependence of the magnetization.[5] In contrast, our measurements reveal a monotonic increase in $\alpha_{xy}^A$ with temperature up to 150K, reaching a maximum value of 1.50 A K⁻¹ m⁻¹, before decreasing to 0.9 A K⁻¹ m⁻¹ at room temperature (Fig. 2(c)). This unconventional temperature dependence of $\alpha_{xy}^A$ indicates that the observed Nernst response cannot be solely attributed to the conventional ANE linked to magnetization scaling up to Curie temperature ($T_c$ = 370K), but rather arises from an additional, distinct Berry curvature driven mechanism dominating transverse thermoelectric transport. A detailed theoretical analysis of the temperature dependence and microscopic origin of $\alpha_{xy}^A$ will be discussed later in this section.

To further clarify the physical origin of the observed large $S_{xy}^A$ value, the total anomalous Nernst power was decomposed to $S_{xy}^A = S_I + S_{II}$, where $S_I = \rho_{XX} * \alpha_{xy}^A$ represents the direct conversion from $\nabla T$ to transverse electric field through the $\alpha_{xy}^A$ and $S_{II} = -S_{xx} * \theta_{AHE}$ corresponds to the Seebeck-anomalous Hall effect mixing term. Notably, $S_I$ reproduces the temperature dependence of the experimentally measured $S_{xy}^A$, whereas $S_{II}$ is small and exhibits no systematic correlation with the overall trend (Fig. 1(f)), indicating that the ANE in $Fe_5Si_3$ predominantly originates from $\alpha_{xy}^A$. Further insight is obtained from the temperature dependence of the anomalous Nernst conductivity normalized by temperature, $\alpha_{yx}^A/T$, shown in Fig. 2(d). Notably, it exhibits a clear $-lnT$ scaling over the temperature range 140-300 K, well below the Curie temperature. Such logarithmic scaling has emerged as a universal hallmark of Berry-curvature-dominated anomalous Nernst transport in magnetic topological materials, including $Co_2MnGa$, $Fe_3Ga$, $YbMnBi_2$, $Co_{3-x}Fe_xSn_2S_2$ and $CoMnSb$[10,20–24]. $Fe_5Si_3$ follows the same trend, as highlighted by the green dotted lines in Fig. 2(d). The robustness of the $-lnT$ behaviour over a wide temperature window implies a broad distribution of Berry curvature near the Fermi level. Consistently, the inset of Fig. 2(d) reveals a linear dependence of $\alpha_{yx}^A/T$ on $lnT$ with a slope of 1.03, in excellent agreement with intrinsic anomalous thermoelectric transport.

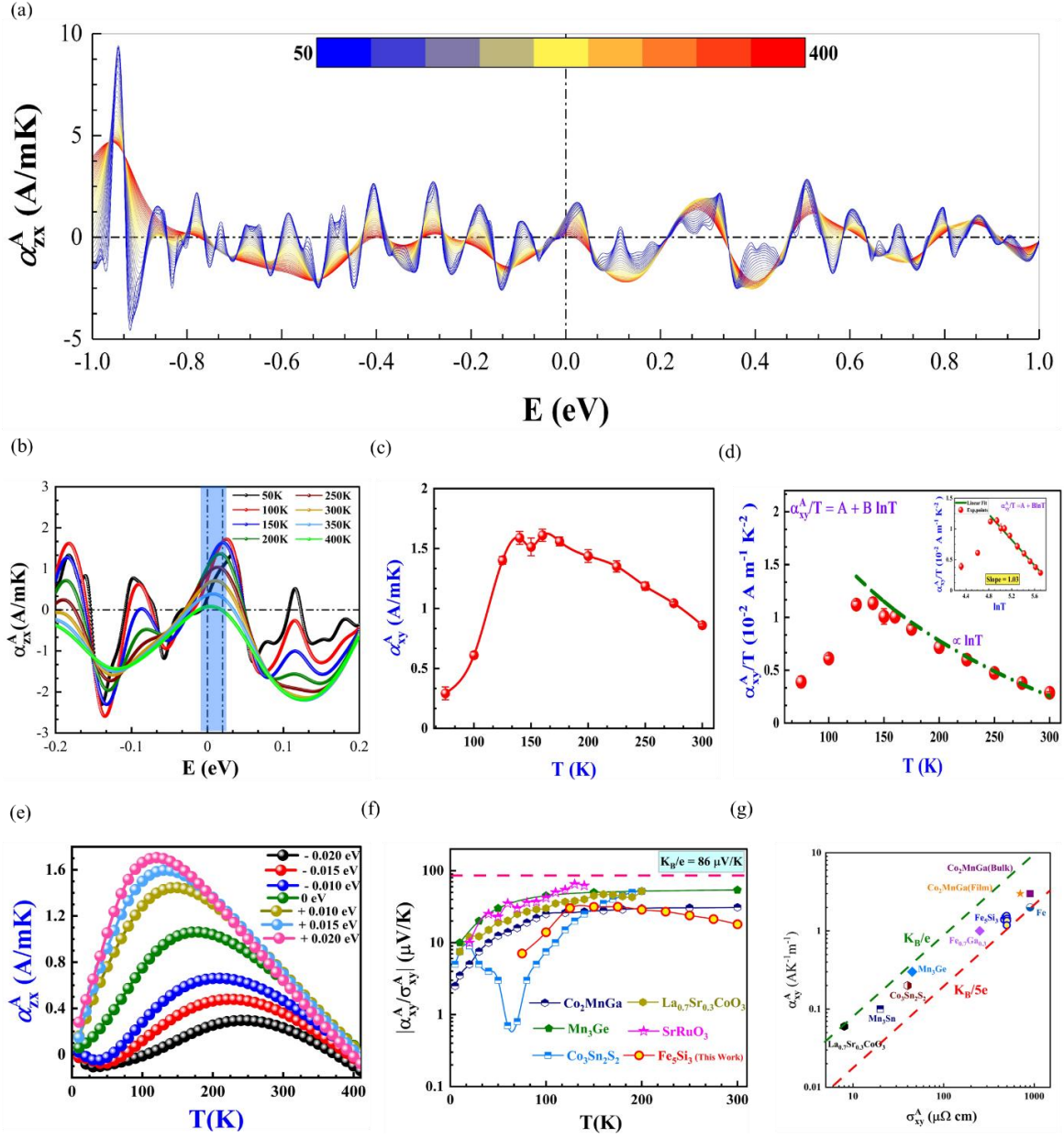

**Figure 2.** (a) ANC of Fe$_5$Si$_3$ as a function of energy at different temperatures. (b) Expanded view of ANC vs E. (c) The temperature dependence of $\alpha_{xy}^A$ (d) The observed $T\ln T$ behaviour of Fe$_5$Si$_3$. (e) Temperature dependence of the calculated ANC with different energy levels (-0.02 eV to 0.02 eV). (f) The $\frac{\alpha_{xy}^A}{\sigma_{xy}^A}$ ratio as a function of temperature in different state-of-the-art topological magnets, along with Fe$_5$Si$_3$. $\frac{K_B}{e}$ is represented with a pink line. (g) $\alpha_{xy}^A$ vs $\sigma_{xy}^A$ of different topological magnets, along with Fe$_5$Si$_3$. The values are well lies between $\frac{K_B}{e}$ (green line) and $\frac{K_B}{5e}$ (red line) ranges. The data of graphs (f) and (g) are taken from the reference [16]

From an intrinsic-mechanism standpoint, the AHE and ANE originate from Berry-curvature-induced transverse carrier motion driven by an electric field and a thermal gradient, respectively. In the low-temperature limit, these two responses are linked by the Mott relation,

$$\frac{\alpha_{xy}^A}{T} = -\frac{\pi^2}{3}\frac{K_B^2}{e}\frac{d\sigma_{xy}^A}{dE} \qquad (1)$$

which implies that an enhanced $\frac{\alpha_{xy}^A}{T}$ reflects a strong energy dependence of $\sigma_{xy}^A$ in the vicinity of the Fermi level. Such sensitivity naturally arises in Weyl and nodal-line systems, where Berry-curvature monopoles and gapped nodal manifolds generate sharply varying $\sigma_{xy}^A(E)$. Beyond the zero-temperature limit, the correspondence between $\alpha_{xy}^A$ and $\sigma_{xy}^A$ is not straightforward and modified by thermal broadening; nevertheless, the ratio $\frac{\alpha_{xy}^A}{\sigma_{xy}^A}$ is expected to remain bounded by the natural scale of $\frac{K_B}{e} = 86\,\mu V\,K^{-1}$, reflecting the fundamental distinction between entropy and charge transport, as argued by Behnia et al[25]. Here, Figures 2(f) and 2(g) reveal a clear correspondence between $\alpha_{xy}^A$ and $\sigma_{xy}^A$ of $Fe_5Si_3$ across a wide temperature range. The scaling behaviour of the epitaxial $Fe_5Si_3$ film closely follows the universal trend reported for intrinsic anomalous transport and remains within the characteristic limits. This observation strongly suggests that the AHE and ANE in $Fe_5Si_3$ are dominated by an intrinsic Berry-curvature-driven mechanism[25,26].

Figure 3(b) presents a logarithmic comparison of the anomalous Nernst coefficient as a function of magnetization across a wide range of magnetic materials. In conventional ferromagnets, the ANE exhibits an approximately linear scaling with magnetization and remains confined to the purple-shaded region, whereas Topological magnet displays a strongly enhanced ANE that markedly exceeds this empirical scaling, forming the green-shaded region. The ANE of $Fe_5Si_3$ lies within this enhanced regime and is comparable to values reported for such materials. Additionally, the anomalous Nernst angle (ANA) $tan\theta_{ANE} = \frac{S_{xy}^A}{S_{xx}}$ shows a pronounced enhancement, reaching a giant value of 0.56 at 175 K and remaining sizable at ~0.25 even at room temperature (Fig. 3(a)); these values are comparable to, and in some cases exceed, the benchmark ANA reported for topological magnetic materials, indicating highly efficient and thermally robust transverse thermoelectric conversion relevant for practical thermoelectric and spin-caloritronic applications.

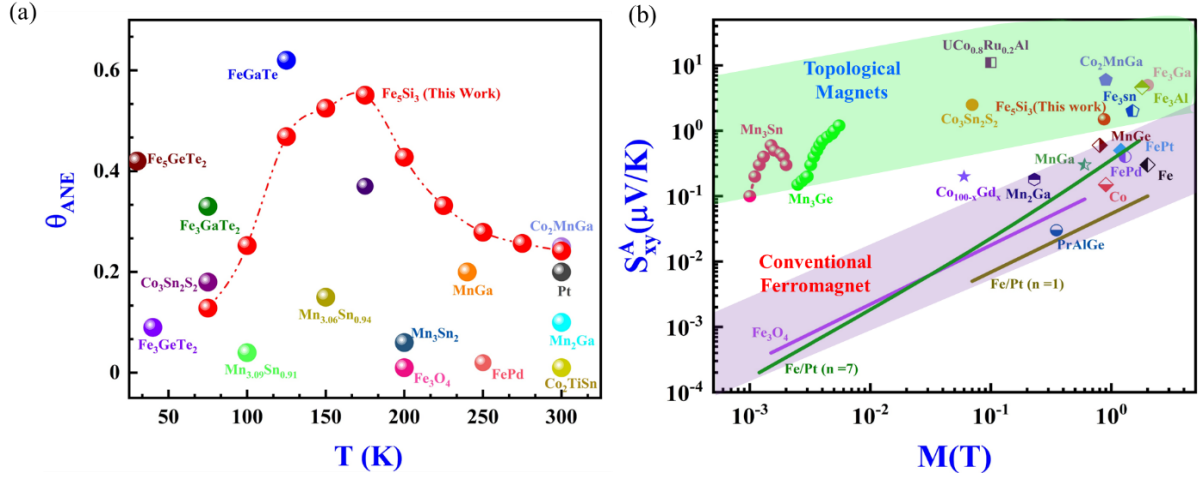

**Figure 3.** (a) Anomalous Nernst angle of state-of-the-art topological materials. (b) Magnetization scaling plot for the maximum values of the ANE in conventional as well as topological magnets.

## **Theoretical Framework of ANC**

To uncover the mechanism underlying the anomalous Nernst response, we analyzed the topological characteristics of the electronic band structure in the vicinity of the Fermi level and evaluate the ANC of $Fe_5Si_3$ using DFT calculations and Wannier tight binding Hamiltonian formalism. A comprehensive analysis of the anomalous Hall conductivity (AHC), including its Berry-curvature origin, has been reported in our previous work.[16] Here, we build directly on those results by employing the calculated AHC as an input, exploiting the fact that the ANC is governed by the energy derivative of the AHC near the Fermi level. ANC can be expressed as

$$\alpha_{zx}(T,\mu) = -\frac{1}{e}\int \frac{\partial f\,(\epsilon-\mu,T)}{\partial \epsilon}\,\frac{\epsilon-\mu}{T}\,\sigma_{zx}(\epsilon)\,d\epsilon \qquad (2)$$

where $\mu$ is the chemical potential. In the low temperature limit, it simplifies to

$$\alpha_{zx} = -\frac{\pi^2 k_B^2 T}{3e}\frac{d\sigma_{zx}}{d\mu} \qquad (3)$$

As illustrated in Fig. 2(a), the $\alpha_{zx}$ exhibits a pronounced dependence on the chemical potential at different temperatures. Notably, $\alpha_{zx}$ attains a peak value of approximately 9.5 A m$^{-1}$ K$^{-1}$ at E = 0.95 eV, which is comparable to those reported in other Weyl semimetals, highlighting the robust thermoelectric response inherent to the system.[5,10,14,27–29] A magnified view shown in Fig. 2(b) further reveals that $\alpha_{zx}$ reaches a value of ~1.5 A m$^{-1}$ K$^{-1}$ around E = 0.015 eV, in close correspondence with our experimentally obtained result. This quantitative agreement between theoretical and experimental data underscores the reliability of the model and affirms the intrinsic nature of the observed thermoelectric behaviour.

To understand the origin of the ANC and its connection to the underlying band topology, we carry out a detailed analysis of the intrinsic electronic structure. We observe that the doubly degenerate bands form a two-dimensional nodal surface along the $k$-path A-L-H-A within the $k_z = \pi$ plane, as illustrated in the red shaded region of Fig. 4 (a). A defining feature of a nodal surface (*NS*) is the linear dispersion of bands perpendicular to the surface (here, along $k_z$). To confirm this, we computed the band structures of $Fe_5Si_3$ along the symmetry directions Γ-A-Γ', M-L-M', and K-H-K', which are perpendicular to the $k_z = \pi$ plane (Fig. 4 (b)). These paths are indicated in the Brillouin zone in Fig. S3(b) of SI, where Γ', K', and M' represent the high-symmetry points projected onto the 001 surface. At the A, L, and H points, all band crossings exhibit twofold degeneracy with linear dispersion, as shown in Fig. 4 (b), confirming the existence of nodal surfaces in the $k_z = \pi$ plane. The detailed symmetry analysis of the nodal surface is discussed below.

In the absence of spin-orbit coupling (SOC), each spin channel behaves effectively as a spinless system because spin and orbital degrees of freedom are independent, which preserves time-reversal symmetry ($\mathcal{T}$). The *NS* arises from the combination of nonsymmorphic $S_{2z}$ symmetry and $\mathcal{T}$. In momentum space, the $S_{2z}$ operation inverts $k_x$ and $k_y$ while leaving $k_z$ unchanged. Without SOC, we have $(S_{2z})^2 = T_{001} = e^{-ik_z}$, where $T_{001}$ represents a fractional translation along the z-direction. The operator $\mathcal{T}$ is antiunitary and inverts $k$ with $\mathcal{T}^2 = 1$. Therefore, the combination $\mathcal{T}S_{2z}$ is antiunitary and only inverts $k_z$. Since $[\mathcal{T}, S_{2z}] = 0$, we obtain $(\mathcal{T}S_{2z})^2 = e^{-ik_z}$. In the $k_z = \pi$ plane, each $k$-point is invariant under $\mathcal{T}S_{2z}$ and $(\mathcal{T}S_{2z})^2 = -1$ leads to a Kramers-like degeneracy. Consequently, the bands along the $k_z = \pi$ plane are doubly degenerate, forming the nodal surface. Including SOC lifts the band degeneracy, thereby breaking the nodal surface. A schematic illustration of the three-dimensional broken nodal surface in the presence of SOC is shown in Fig. 4 (e).

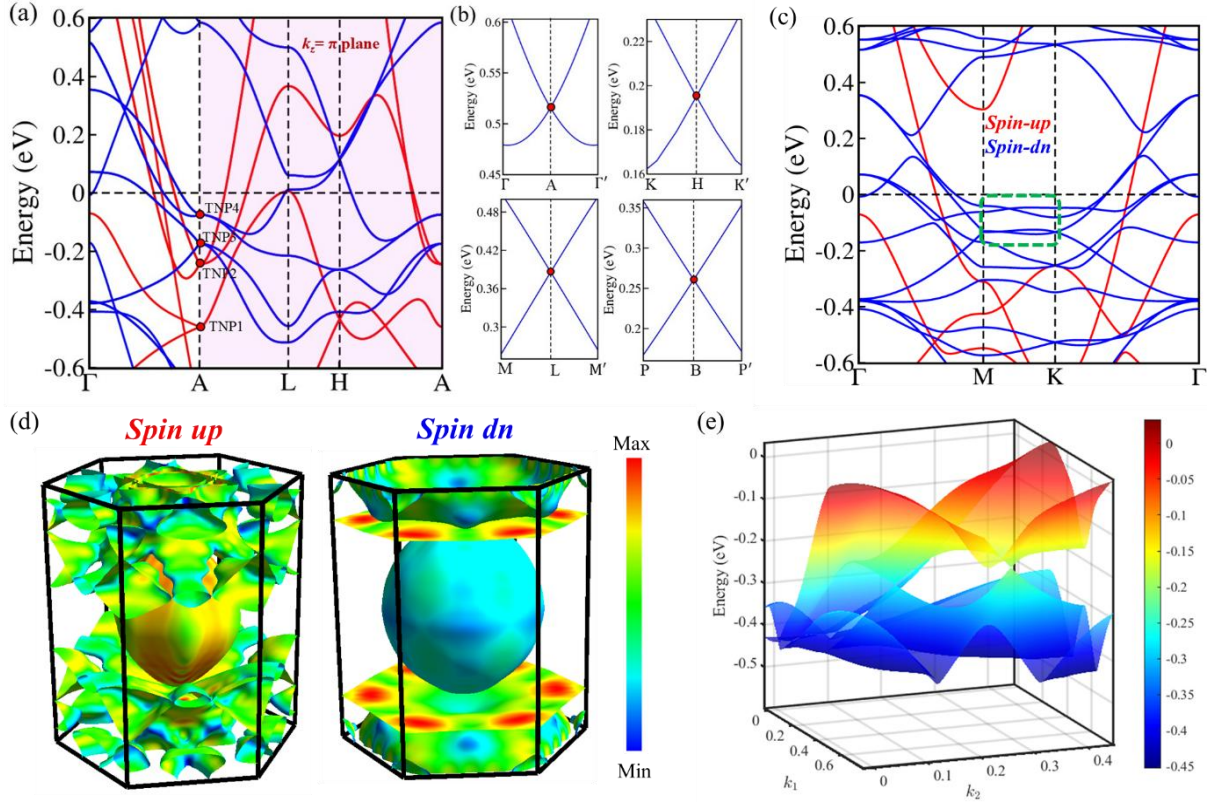

**Figure 4.** (a) Electronic band structure along the Γ-A-L-H-A path. (b) Continuous band crossing along the k path perpendicular to the $k_z = \pi$ plane without SOC. The bands at A, H and L exhibit two-fold degeneracy with linear dispersion. (c) Band Structure along Γ-M-K-Γ path with green dashed rectangle represents flat band dispersion. (d) Fermi surface of spin up and spin down, overlaid with a colour map indicating Fermi velocity. (e) 3D band dispersion of nodal surface with SOC. The energy band gap opens on the nodal plane after introducing SOC.

The Fermi surface, shown in Fig. 4 (d), consists of several interesting Fermi pockets composed of both electron and hole charge carriers, which are discussed in detail in the section 3 of SI. Here, we focus specifically on the two flat one-dimensional sheet-like Fermi surfaces present in spin down channels as shown in Fig. 4(d). These features originate from two flat bands crossing the Fermi level along the Γ–M–K–Γ path as shown in Fig. 4 (c). These bands subsequently merge and form a nodal surface in the $k_z = \pi$ plane, and cross in the $k_z = 0$ plane, forming a nodal line. Such flat bands are typically associated with Kagome lattices; however, in the present case, they emerge in a triangular lattice structure. Similar to nodal lines, nodal planes are also ubiquitous in electronic band structures in the absence of SOC. A nodal plane generally reduces to nodal lines upon taking a constant energy cut, whereas a nodal line reduces to discrete nodal points. Beyond this general property, the flat band dispersion discussed in this work lies close to a stationary point of the nodal plane, which can significantly enhance the density of states and thereby amplify the intrinsic thermoelectric properties. We emphasize that the sharp peaks in the Berry curvature do not originate solely from the broken nodal lines (or nodal rings), as discussed earlier in our earlier study, but also from the lifting of degeneracy within the nodal surface, which leads to the formation of nodal points. The combined effects of broken nodal lines and emergent nodal surfaces enhance berry curvature, which is likely responsible for the observed enhancement of the ANE.

**Topological Nernst effect**

In addition to the large ANE and ANA, we also observe a sizable topological Nernst effect (TNE). Similar to the Topological Hall effect (THE), the TNE represents a distinct thermoelectric response driven by real-space Berry curvature originating from topologically nontrivial spin textures, generating an emergent transverse electric field under a thermal gradient.[30–32] Our previous detailed investigation of THE in the same material suggests the possibility of nontrivial spin textures[16], indicating a strong likelihood of TNE emerging from similar topological spin configurations in this system. The TNE component can be extracted from total Nernst effect by using the relation, $S_{xy} = S_0 + S_{xy}^A + S_{xy}^T$, where $S_0$ ($\propto H$), $S_{xy}^A$ ($\propto M$) and $S_{xy}^T$ are ordinary, anomalous and topological Nernst components respectively (see Section 2 of SI for details). By isolating $S_{xy}^T$ in the temperature range of 150–325 K, we find that the TNE exhibits an almost linear temperature dependence (Fig. 5(b)), reaching a maximum value of ~0.43 µV K$^{-1}$ at 325 K (Fig. 5(a)), which is comparable to those reported for related topological magnets.[12,33–36] Notably, both THE and TNE signals display a pronounced maximum near 0.3T, suggesting the persistence of chiral spin-textures and reinforcing the robustness of topological transport phenomena across different measurement configurations.

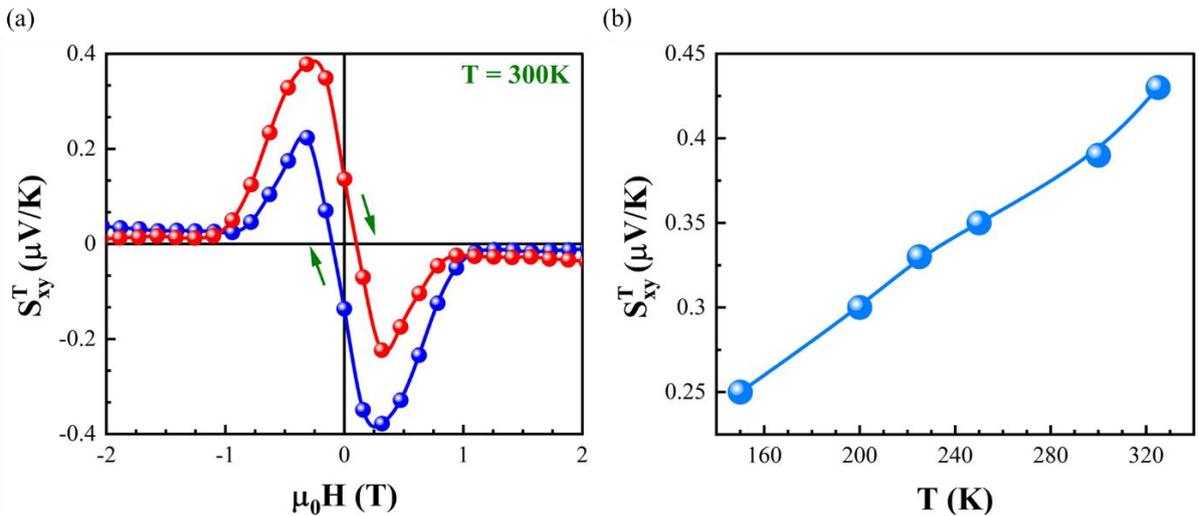

**Figure 5.** (a) Topological Nernst effect of Fe$_5$Si$_3$ epitaxial thin film at 300K. (b) Temperature dependence of the maximum value of TNE.

## Conclusion

In this study, transverse thermomagnetic transport in $Fe_5Si_3$ is shown to be dominated by a large anomalous Nernst response, with a transverse Nernst signal reaching 1.50 µV K$^{-1}$ and an anomalous Nernst angle of ≈ 0.56, indicating highly efficient transverse thermoelectric conversion and directly highlighting the suitability of this material for compact, single-material based transverse thermoelectric device architectures. Beyond the anomalous contribution, a substantial topological Nernst contribution of ~ 0.43 µV K$^{-1}$ persisting above room temperature indicates the presence of additional involvement of real-space Berry curvature from nontrivial spin-textures. Quantitative agreement between experiment and first-principles DFT calculations shows that this behaviour is not governed by conventional magnetization scaling but instead arises from a sharply energy-dependent Berry curvature generated by the coexistence of Weyl nodal lines, nodal surfaces, and nearly flat bands near the Fermi level, which together control both the magnitude and temperature evolution of the Nernst response. These findings establish $Fe_5Si_3$ as a model rare earth free binary topological magnet in which multiple band-topological features act cooperatively to enhance thermomagnetic transport, highlighting the broader potential of Topological ferromagnetic compounds for Berry-curvature-driven transverse thermoelectric functionalities.

## <u>Acknowledgments</u>

S.P. acknowledges the University Grant Commission, Government of India, for financial assistance. The authors acknowledge the Institute's National Research Facility for EDAX measurements, the Central Research Facility for PPMS and XPS measurements, and the Physics Department Facilities for MPMS and XRD measurements. The authors gratefully acknowledge the Indian Institute of Technology Hyderabad and the National Supercomputing Mission (NSM) for providing computational resources through 'PARAM SEVA' at IIT Hyderabad, implemented by CDAC. S.P. acknowledges financial support from DST-INSPIRE through a research fellowship. V.K. acknowledges support from the DRDO project (ERIP/ER/202312003/M/01/1853). The authors thank N. Kojima for technical support with thermoelectric measurements and device fabrication. The authors acknowledge the National Institute for Materials Science (NIMS) for providing research facilities and technical support. Financial support from JST ERATO "Magnetic Thermal Management Materials" (Grant No. JPMJER2201) and JST CREST "Exploring Innovative Materials in Unknown Search Space" (Grant No. JPMJCR21O1) is also acknowledged.

# Supplementary Information

# Nodal-Surface and Flat-Band Driven Large Anomalous Nernst effect in Epitaxial Ferromagnetic Weyl Metal Fe$_5$Si$_3$

Shubhashish Pati[1], Sonali Srotaswini Pradhan[2], Nanhe Kumar Gupta[3], Abhay Pandey[1], Nikita Sharma[1], Nakul Kumar[1], Saurav Singh[1], Yuya Sakuraba[3], V. Kanchana[2]*, Sujeet Chaudhary[1]*

[1] Thin Film Laboratory, Department of Physics, Indian Institute of Technology Delhi, New Delhi 110016, India

[2] Department of Physics, Indian Institute of Technology Hyderabad, Kandi 502285, Sangareddy, Telangana, India

[3] Centre for Magnetic and Spintronic Materials (CMSM), National Institute for Materials Science (NIMS) 1-2-1 Sengen, Tsukuba, Ibaraki 305-0047, Japan

**Keywords**: *Sputtered Epitaxial film, Anomalous Nernst effect, Giant Anomalous Nernst angle, Nodal surface, Flat band, Magnetic Weyl, Topological Nernst effect.*

Corresponding Authors

E-mail address: sujeetc@physics.iitd.ac.in (S. Chaudhary), kanchana@phy.iith.ac.in (V. Kanchana)

# 1. Computational details

All structural relaxations were carried out within the framework of density functional theory using the Vienna *ab initio* Simulation Package (VASP). The exchange-correlation interaction was treated using the Perdew-Burke-Ernzerhof (PBE) functional formulated within the generalized gradient approximation. To account for the on-site Coulomb interaction of the Fe 3d electrons, calculations were performed within the GGA+U scheme, adopting an effective Hubbard parameter $U = 0.45$ eV, consistent with prior reports. A plane-wave basis with a kinetic energy cutoff of 600 eV was employed for all computations, and total energies were converged to within $10^{-8}$ eV. Brillouin-zone integrations were performed using an $8 \times 8 \times 10$ Monkhorst-Pack k-point mesh. To explore the topological characteristics of the electronic structure, a tight-binding model was constructed based on maximally localized Wannier functions. The surface electronic states and associated topological features were subsequently evaluated using the iterative Green's function technique as implemented in the *Wannier Tools* package.

# 2. Thermal and Electrical Transport Measurements and Device Fabrication

Electrical and thermoelectric transport measurements, including longitudinal resistivity ($\rho_{xx}$), anomalous Hall effect (AHE), Seebeck effect (SE), and anomalous Nernst effect (ANE), were performed on $Fe_5Si_3$ films grown on GaAs substrates with dimensions of $10 \times 7$ mm$^2$ (Fig. 1(a)). The device structures were defined using standard photolithography followed by Ar-ion milling. For the measurements of $\rho_{xx}$, $\rho_{yx}$, and $S_{SE}$, a Hall bar geometry with a channel width of 200 μm and length of 5 mm was employed. For ANE measurements, the $Fe_5Si_3$ film was patterned into an array of 30 parallel wires, each 60 μm wide and 2 mm long, connected by Au leads to form a meander structure in order to enhance the output voltage. An on-chip Au-wire thermometer was fabricated adjacent to the Hall bar and used to determine the temperature gradient ($\nabla T$) induced by the heater at each measurement temperature.

The Hall voltage was measured using a Physical Property Measurement System (PPMS DynaCool; Quantum Design) by sweeping the external magnetic field H up to 3 T, applied perpendicular to the substrate surface, under a constant current of 100 μA. The Hall resistivity was calculated from $\rho_{yx} = R_{yx}*t$, where $R_{yx}$ is the measured Hall resistance and t is the thickness of the film, over a temperature range of 75-300K. The SE and ANE were also measured in the temperature range of 75-325 K. The experimental setup and the corresponding sample holder used for SE and ANE measurements are shown in Fig. 1(a). During the measurements, the magnetic field was applied perpendicular to the film surface (along the z-axis), while a temperature gradient ∇T was applied along the x-axis. The samples were mounted on a custom-built holder equipped with an integrated chip heater. The magnitude of ∇T was controlled by adjusting the voltage applied to the heater and was determined using on-chip Au thermometers fabricated on the film surface. The voltages generated by the SE and ANE were simultaneously recorded for the Hall bar and meander-patterned devices, respectively, using nanovoltmeters. All SE and ANE measurements were conducted using a Versa Lab system (Quantum Design).

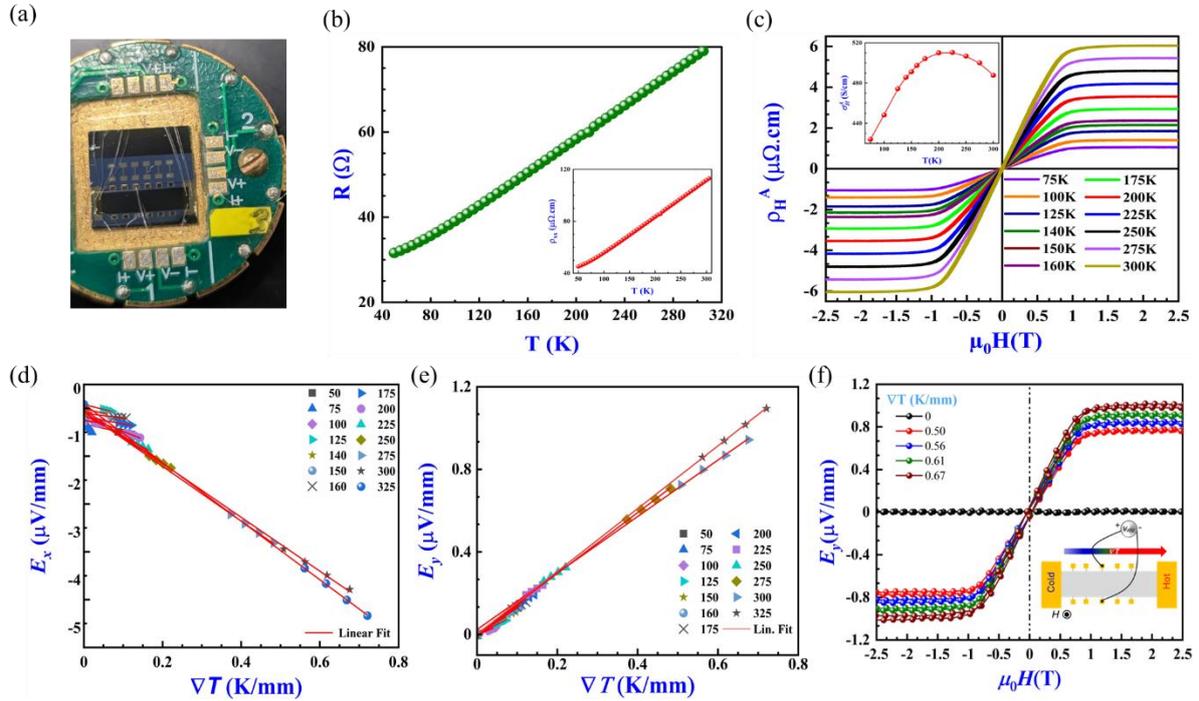

**Figure S1.** (a) Real image of patterned $Fe_5Si_3$ thin film and sample holder of AHE measurements. (b) Temperature dependence of Longitudinal resistance (inset shows longitudinal resistivity value). (c) Anomalous hall resistivity and anomalous hall conductivity (inset) at different temperature ranges. (d) Temperature-dependent Seebeck effect-induced $E_x$ measured for different ∇T. (e) Linear fit of $E_y$ versus ∇T for temperatures ranging from 75 K to 325K K. (f) Magnetic field-dependent ANE-induced $E_y$ measured at 300 K for different ∇T.

## 3. Topological Nernst effect

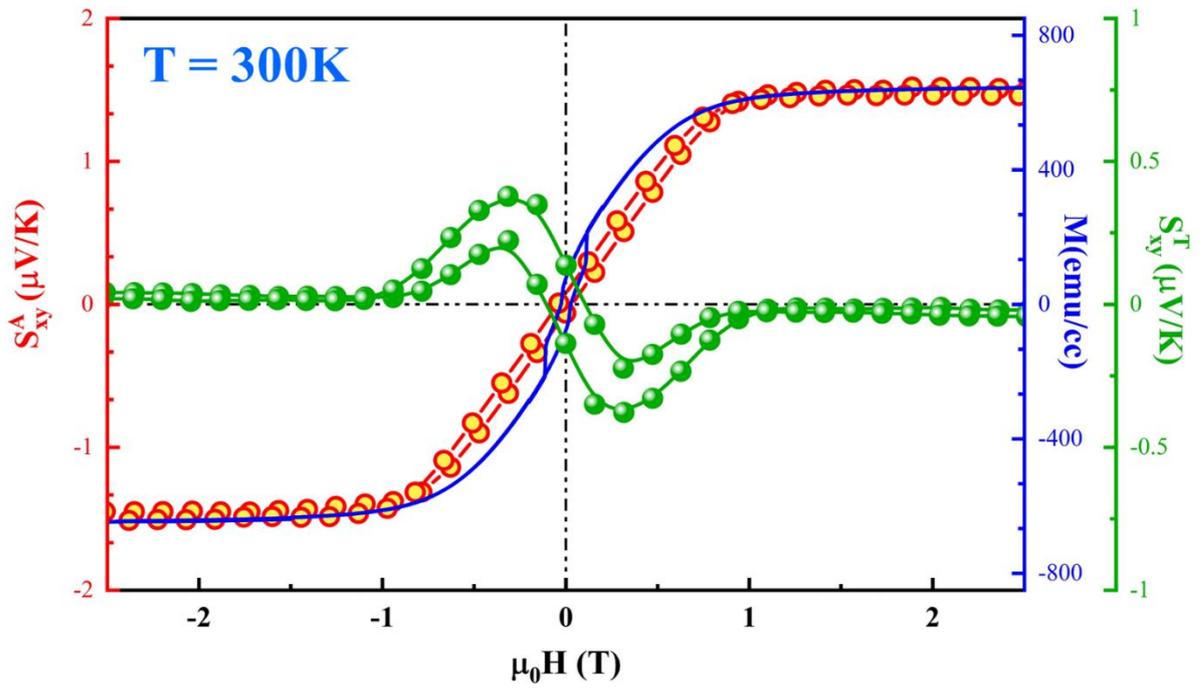

**Figure S2**. Magnetic field dependence of total Nernst effect coefficient $S_{xy}^A$ and the magnetization curves at 300K. Green curves represents the extracted Topological Nernst coefficients $S_{xy}^T$.

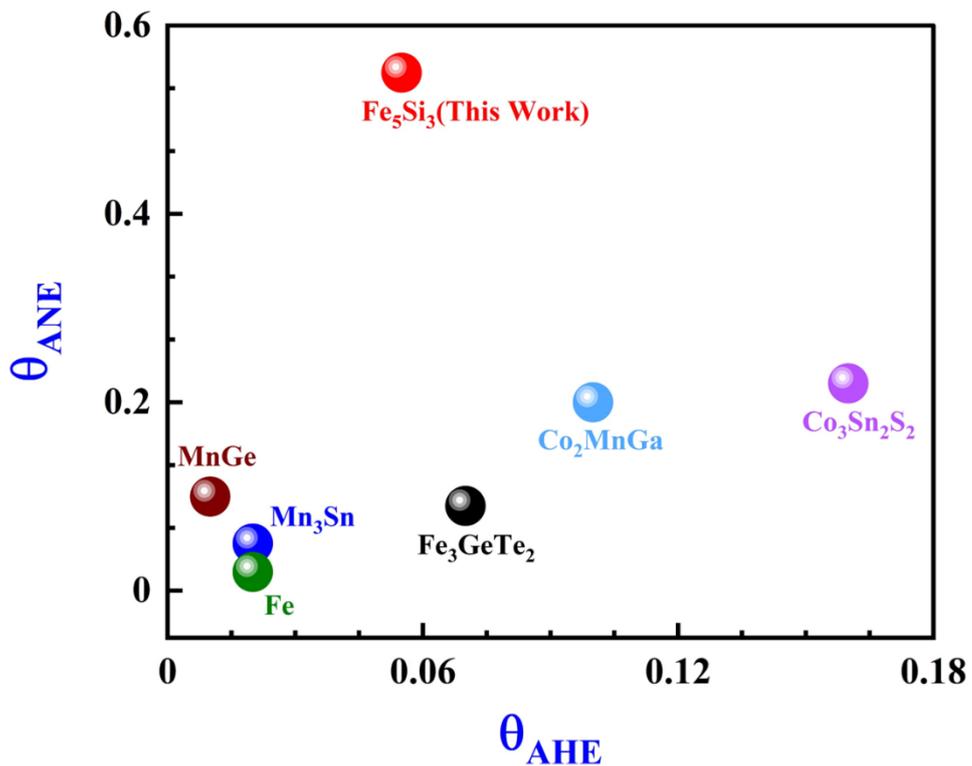

**Figure S3**. Anomalous hall angle and anomalous Nernst angle of different topological materials.

## 4. Fermi Surface Topology

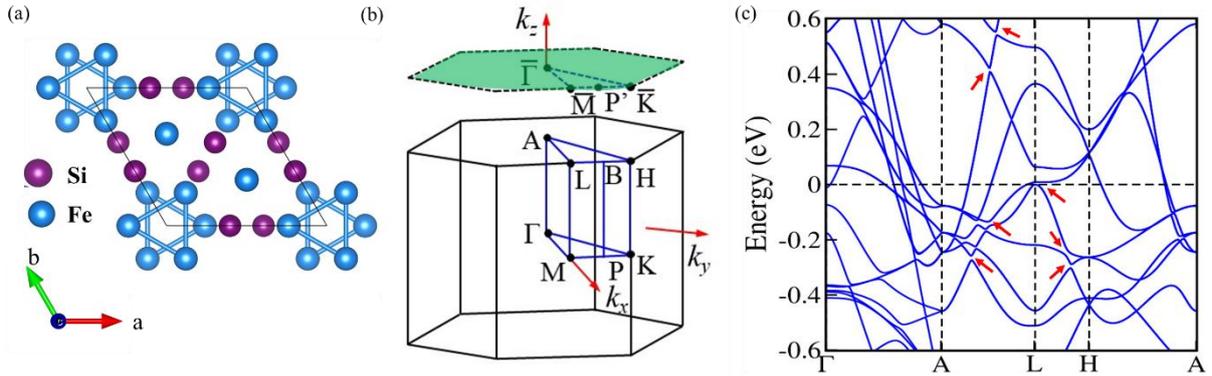

**Figure S3**. (a) The crystal structure of $Fe_5Si_3$, (b) Bulk BZ and its projected (001) surface, depicted by the green-shaded region represents nodal surface, (c) Band structure with SOC, with red arrow indicate the gapped crossing points in nodal surface region.

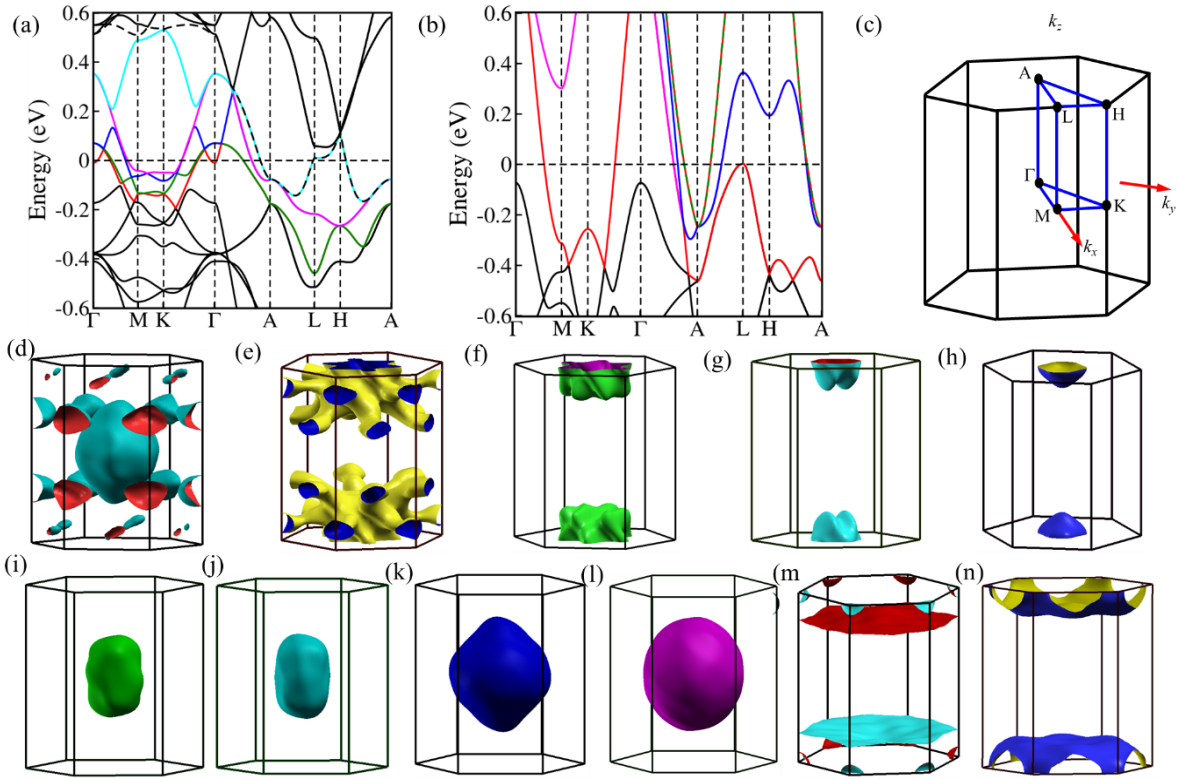

**Figure S4**. (a, b) The electronic structure for spin-down and spin-up states, respectively, without spin-orbit coupling, with coloured bands intersecting the Fermi level. (c) The high symmetry paths (blue coloured) of the BZ. (d-n) Fermi pockets corresponding to coloured band crossing the Fermi level (as shown in Fig. a and b)

The Fermi surface plot for the bands crossing the Fermi level ($E_F$) in $Fe_5Si_3$ without SOC, shown in Fig. 4(d-n), reveals eleven distinct pockets along the high-symmetry path Γ-M-K-Γ-A-L-H-A, corresponding to four different bands crossing $E_F$. Among these, the first five pockets correspond to spin-up bands crossing the Fermi level, while the remaining six correspond to spin-down bands. Closed Fermi surfaces are formed by bands crossing in a distinct plane, and the last two flat 1D sheet-type Fermi surfaces arise from flat bands along the M-K path, marked by red and green bands. As these bands cross $E_F$ twice, transitioning from the valence to the conduction band and vice versa, the system exhibits both electron- and hole-like charge carriers.